\documentclass[reprint, pra, twocolumn,showpacs,preprintnumbers,amsmath,amssymb]{revtex4-2}

\usepackage{graphicx}
\usepackage{natbib}
\usepackage{dcolumn}
\usepackage{bm}
\usepackage[colorlinks=true,linkcolor=blue, citecolor=blue, allcolors=blue]{hyperref}
\usepackage{verbatim} 
\usepackage{units}
\usepackage{tipa}
\usepackage{physics}
\usepackage{booktabs}
\usepackage{subfigure}
\usepackage{multirow}
\usepackage{soul}
\newcolumntype{C}{>{$\displaystyle}c<{$}}

\usepackage{xcolor}
\newcommand{\Rb}[1]{$^{#1}$Rb}
\newcommand{\Emax}{E_\mathrm{max}}

\newcommand{\sigmav}{\langle \sigma_\mathrm{tot} v\rangle}

\newcommand{\svl}{\langle \sigma_{\rm{loss}}  v \rangle}
\newcommand{\svlu}{\langle  \sigma_{\rm{loss}} (U)\ v \rangle}

\newcommand{\svt}{\langle \sigma_{\rm{tot}} v \rangle}

\newcommand{\gloss}{\Gamma_{\rm{loss}} }
\newcommand{\gtot}{\Gamma_{\rm{tot}} }

\newcommand{\PQDU}{p_{\rm{QDU6}}}

\newcommand{\ud}{U_{\rm{d}}}

\newcommand{\kb}{k_{\rm{B}}}

\newcommand{\bea}{\begin{eqnarray}}
\newcommand{\eea}{\end{eqnarray}}

\newcommand{\ma}{m_{\mathrm{t}}}
\newcommand{\mb}{m_{\mathrm{bg}}}

\begin{document}

\title{Measurement of Rb-Rb van der Waals coefficient via\\Quantum Diffractive Universality}


\author{Riley A. Stewart$^{1*}$, Pinrui Shen$^1$, James L. Booth$^2$, Kirk W. Madison$^1$}
\affiliation{$^1$Department of Physics \& Astronomy, University of British Columbia,\\
6224 Agricultural Road, Vancouver, BC, V6T 1Z1, Canada}

\affiliation{$^2$ Physics Department, British Columbia Institute of Technology, \\3700 Willingdon Avenue, Burnaby, B.C. V5G 3H2}

\address{$^*$Corresponding author: rstewart@phas.ubc.ca}



\begin{abstract}
Collisions between trapped atoms or trapped molecules with room temperature particles in the surrounding vacuum induce loss of the trapped population at a rate proportional to the density of the background gas particles.  The total velocity-averaged loss rate coefficient $\sigmav$ for such collisions and the variation of the loss rate with trap depth has been shown to depend only on the long range interaction potential between the collision partners \cite{booth2019universality}.  This collision universality was previously used to realize a self-calibrating, atom-based, primary pressure standard and was validated by indirect comparison with an orifice flow standard \cite{shen2020realization}.  Here, we use collision universality to measure  $\sigmav = 6.44(11)(5) \times 10^{-15}~\rm{m^3/s}$ for Rb-Rb collisions and deduce the corresponding $C_6 = 4688(198)(95)~E_ha _0^6$, in excellent agreement with predictions based upon \textit{ab initio} calculated and previously measured $C_6$ values \cite{derevianko2001, c6atominter, c6collisionspectro}.
\end{abstract}


\maketitle



Since the pioneering work of Rutherford \cite{Rutherford1899}, the study of collisions between particles has been central to the advancement of our knowledge of atomic, molecular, and particle physics \cite{child1996molecular}.  Recently, ultra-cold atom-atom collisions have been used provide precision values of the leading term of the van der Waals attractions through the measurement of magnetically tunable Feshbach resonances \cite{RevModPhys.82.1225,Chin2004,vanKempen2002}.  One of the first such measurements provided the Rb-Rb $C_6$ coefficient \cite{c6collisionspectro}, in agreement with later measurements of the Rb ground-state polarizability using atom interferometry \cite{c6atominter}.  Here, we present another experimental method of determining the leading term of the van der Waals interaction.  Namely, we determine it from measurements of the energy exchange produced by collision between room-temperature particles and ultra-cold trapped atoms.

In our previous work, Refs.~\cite{booth2019universality,shen2020realization}, we demonstrated that the total collision cross section of and the energy exchanged by room-temperature collisions are universal functions depending only on the leading term of the van der Waals interaction.  In particular, for neutral atom-atom collisions, where the leading term of long range interaction is given at an atom separation $R$ by $V(R) =-C_6/R^6$, the total collision rate is,
\begin{eqnarray}
\gtot = n \svt
\label{eqn:gtot}
\end{eqnarray}
with total velocity-averaged loss rate coefficient,
\begin{eqnarray}
\svt &\approx& 10.45812\left(\frac{C_6}{\hbar} \right)^{\frac{2}{5}} \left(\frac{\kb T}{\mb}\right)^{\frac{3}{10}} \nonumber \\
 & & + 6.71531\left(\frac{C_6}{\hbar} \right)^{\frac{1}{5}} \left(\frac{\mb}{\kb T}\right)^{\frac{1}{10}} \left(\frac{\hbar}{\mu}\right).
 \label{eq:svt_C6}
 \end{eqnarray}
Here $n$ and $\mb$ are the density and mass of the background gas particles and $\mu$ is the reduced mass of the collision pair.  The angle brackets denote the average over the relative collision velocities given the Maxwell-Boltzmann distribution for room temperature ($294$ K). 

If we consider an ensemble of sensor atoms held in a trap of finite depth with zero initial energy, the collision induced sensor ensemble loss rate is
\begin{eqnarray}
\gloss = \gtot \left[1 - \PQDU\right] = n \svl
\label{eqn:gloss}
\end{eqnarray}
where $\PQDU$ is the probability that a sensor atom remains in the trap after a collision (mediated by a $-C_6/R^6$ long range interaction) corresponding to the atom's post-collision cumulative energy distribution evaluated at the trap depth.  As shown in Refs.~\cite{booth2019universality,shen2020realization}, this energy exchange distribution is a universal function and, at small trap depths ($U\leq\ud$), the CDF is well-modeled by
\begin{eqnarray}
\PQDU = \sum_{j=1}^{\infty}\beta_j\left(\frac{U}{\ud}\right)^j
\label{eq:svl_universal}
\end{eqnarray}
where the coefficients $\beta_j$ are given in Tab.~\ref{beta} and $\ud$ is the characteristic energy scale for so-called quantum diffractive collisions \cite{bali1999quantum, fagnan2009observation},
\begin{eqnarray}
\ud &=& \frac{4\pi \hbar^2}{\ma \bar{\sigma} }\ = \  \frac{4\pi \hbar^2 v_p}{\ma \svt}.
\label{eqn:Ud}
\end{eqnarray}
Here $\ma$ is the mass of the trapped atom and $v_p=\sqrt{2\kb T/\mb}$ is the most probable velocity of the background particles (of mass $\mb$ at temperature $T$).
\begin{table}[!ht]
\centering
\begin{tabular}{|c|c|}
\hline
Term & $\beta_j$  \\
\hline
1 & 0.673 (7)  \\
2 & -0.477 (3) \\
3 & 0.228 (6)\\
4 & -0.0703 (42)\\
5 & 0.0123 (14)\\
6 & -0.0009 (2)\\
\hline
\end{tabular}
\caption{The universal coefficients appearing in Eq.~\ref{eq:svl_universal} were derived from fitting the quantum scattering (QS) computations for $\svlu$ to a sixth order polynomial.  The derivation of $\beta_j$ can be found in  \cite{booth2019universality}.}
\label{beta}
\end{table}
From Eqs.~\ref{eqn:gtot} -- \ref{eqn:Ud}, we see that the measured loss rate depends on two parameters, $n$ and  $\svt$. Normalizing the measured loss rate (Eq.~\ref{eqn:gloss}) by its value extrapolated to zero trap depth, $ \gloss(U=0) = n \langle \sigma_{\rm{tot}} v \rangle$, eliminates the background density and we obtain,
\begin{eqnarray}
\frac{\Gamma_{\rm{loss}}}{\gloss(U=0)} = \frac{\svl}{\svt}&=& 1 - p_{\rm{QDU6}}.
\label{eqn:gloss_g0}
\end{eqnarray}
Thus, measurements of the loss rate as a function of trap depth, normalized as above, enables the experimental determination of $\ud$ and therefore $\svt$ for any collision partners. This procedure was previously used to realize a self-calibrating, atom-based, primary pressure standard and was validated by comparison with an ionization gauge that was calibrated by the orifice flow standard at National Institute of Technology (NIST) \cite{booth2019universality,shen2020realization,Madison2018}.

In this work, we use the normalized loss rates of $^{87}$Rb atoms held in a magnetic trap exposed to a room-temperature background of Rb atoms (both $^{87}$Rb and $^{85}$Rb atoms) to determine the total loss rate coefficient $\svt$ and corresponding $C_6$ value for Rb-Rb interactions.  After eliminating losses arising from two-body intra-trap collisions and accounting for collision-induced sensor ensemble heating, we find a value for $\svt$ in excellent agreement with predictions based upon \textit{ab initio} calculated \cite{c6collisionspectro} and previously measured $C_6$ values \cite{derevianko2001, c6atominter}.

\section{Measurement Theory}\label{sec:theory}

\subsection{Trap losses}

In order to extract the $C_6$ value for Rb-Rb interactions, we use loss rate measurements of an ultra-cold atomic ensemble of \Rb{87} atoms in a magnetic trap subject to collisions with other Rb atoms in the surrounding background vapour.  However, as atoms held in the magnetic trap are subject to several possible loss channels, we must isolate only those losses induced by Rb-Rb collisions to accomplish this goal.

For example, the total collision rate for a Rb sensor atom in the magnetic trap can be expressed as
\begin{equation}
	\Gamma_\mathrm{tot} = \sum_i n_i \langle \sigma_\mathrm{tot} v \rangle_{\mathrm{Rb}-i}
	\label{eq:summedLoss}
\end{equation}
where the sum is taken over all possible constituents, labelled by $i$, in the background vapour. Here, $n_i$ is the background density of the constituent, and $\langle \sigma_\mathrm{tot} v\rangle_{\rm{Rb-i}}$ is the total loss rate coefficient associated with collisions with the trapped \Rb{87}.  In addition, Majorana spin-flip loss ($\Gamma_\mathrm{maj.}$) can occur as atoms pass through the vicinity of the field zero, where they have a finite probability of transitioning to an untrappable spin state, and are thereby lost from the trap \cite{PhysRevA.74.035401}.  In order to isolate losses induced only by Rb-Rb collisions, we vary the Rb density in the vacuum while keeping the trapping gradient and the background gas partial pressures constant. This insures that  $\Gamma_\mathrm{maj.}$ and losses induced by collisions with other gases do not vary.

Intra-trap collisions between trapped atoms may promote one or both collision partners above the trap depth, inducing loss proportional to the density of the trapped ensemble. The change in the trapped atom number $N$ over holding time $t$ can be written as,
\begin{equation}
	\frac{dN}{dt} = -\left(\sum_i n_i \langle \sigma v\rangle_i  + \Gamma_\mathrm{maj.}\right) N - L_2 \int n^2(\mathbf{r},t) d \mathbf{r}.
	\label{eq:dNdt}
\end{equation}
The two-body loss coefficient $L_2$ includes both elastic \cite{Roberts1998} and inelastic \cite{mies1996estimating} collisional loss channels between trapped \Rb{87} atoms. Assuming a trapped population of $10^6$ atoms and a trapping volume of 1 cm$^3$, an order of magnitude estimate for the inelastic loss rate is $L_{2,\mathrm{in.}} n  N \simeq 10^{-3} $ Hz, well below the loss rate due to collisions with background particles ($\geq 0.1$ Hz in this work), and can therefore be neglected \cite{mies1996estimating}. 

Two-body elastic losses, however, cannot be neglected.  The principle problem is that these two-body losses are trap-depth dependent.  This dependence arises due to the decrease of the trapping volume induced by the resonant radio-frequency (RF) field used experimentally to set the trap depth $E_\mathrm{max}$ (see Sec. \ref{sec:exp}). Atom pairs with combined energies greater than $E_\mathrm{max}$ can exchange energy to promote one partner above $E_\mathrm{max}$, and thus be ejected from the trap. The effective contribution of the elastic two-body loss rate to the total loss rate scales as,
\begin{equation}
	\frac{\int_{\Omega} n^2(\mathbf{r},t) d \mathbf{r}}{\int_{\Omega} n(\mathbf{r} ,t) d \mathbf{r}}
	\label{eq:effectiveintra}
\end{equation}
where $\Omega$ denotes the region in space where the energy of atoms remaining in the trap is below $E_\mathrm{max}$. Owing to the non-uniformity of the spatial distribution in the trap \cite{jurgilas2021collisions}, the effective intra-trap density scales inversely with trapping volume $\Omega$, and hence the effective contribution to the total loss rate via Eq.~\ref{eq:effectiveintra} varies as a function of trap depth $\Emax$. As the total collision rate coefficient $\sigmav$ is uniquely determined by the variation of the loss rate with trap depth $\Emax$, it is imperative to minimize or eliminate these losses to determine $\sigmav$ accurately.

To suppress elastic two-body losses, we apply an RF field to the atoms initially loaded into the magnetic trap to eject all atoms above a defined `cut' energy $E_\mathrm{cut} = \Emax/2 $.  This depletes the population of atom pairs with a combined energy above $\Emax$, and hence elastic two-body losses are energetically forbidden until significant heating of the ensemble occurs.  Both collisions with the background particles and multiple intra-trap collisions can populate states above $E_\mathrm{cut}$. We mitigate these effects by restricting the holding time to less than 1.2 lifetimes.

In the absence of two-body losses, Eq.~\ref{eq:dNdt} can be solved to find,
\begin{equation}
	N(t) = N_0 e^{-(\Gamma_\mathrm{Rb} + \Gamma_0) t}
	\label{eq:onebody}
\end{equation}
where $\Gamma_\mathrm{Rb}$ denotes the loss rate due to collisions with background rubidium, and $\Gamma_0$ is the aggregate loss rate due to collisions with all other background species and Majorana spin-flips.

\subsection{Ensemble distribution and Heating}

For an atom with finite energy $E$, the effective trap depth experienced is given by the difference $(E_\mathrm{max} - E)$, the energy required to exceed $E_\mathrm{max}$. Consequently, the loss rate from the trap is given by the loss rate coefficient $\langle \sigma_\mathrm{loss} v \rangle$ (appearing in Eq.~\ref{eqn:gloss} and Eq.~\ref{eqn:gloss_g0}) averaged over the energy distribution of the trapped ensemble $\rho(E)$,
\begin{equation}
	\overline{\langle \sigma_\mathrm{loss}(\Emax) v \rangle} = \frac{\int_0^{\Emax} \rho(E) \langle\sigma_\mathrm{loss} (\Emax - E) v \rangle dE}{\int_0^{\Emax} \rho(E) dE}.
	\label{eq:intlosscoeff}
\end{equation}
The energy distribution can be measured experimentally by varying the trap depth $\Emax$ and observing the remaining population. This corresponds to a direct measurement of the cumulative energy density function and therefore $\rho(E)$, as described previously \cite{booth2019universality, shen2020realization}.

\begin{figure*}[t!]
    \begin{center}
    \includegraphics[width=1\textwidth]{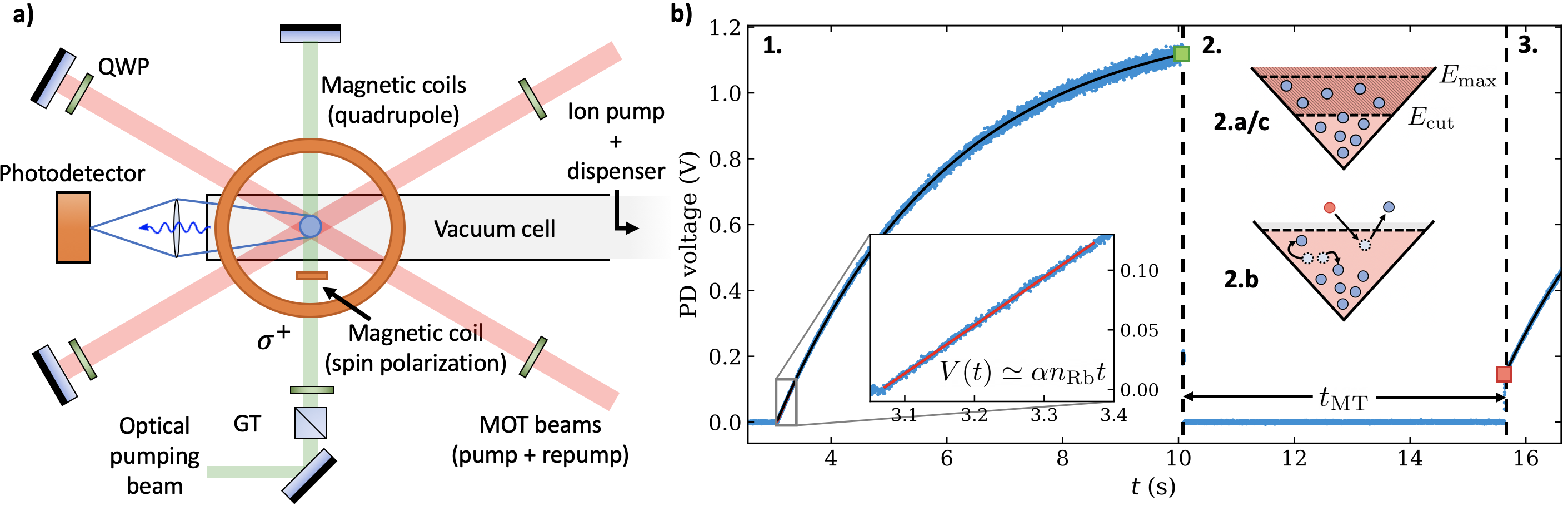}
    \caption{Experimental apparatus and procedure. Panel (a) depicts a schematic of the experimental setup surrounding the vacuum cell and the relative orientation of the trapping light and magnetic fields. The pump and repump light are combined through polarized beam splitters prior to the cell. A Glan-Thompson (GT) maintains the polarization of the optical pumping beam. An additional magnetic field is generated by a small coil to ensure the optical pumping beam drives the required sample-polarizing $\sigma^+$ transitions. Panel (b) shows the atomic fluorescence, as monitored by the photodetector shown in panel (a), for a single shot. In stage (1), rubidium atoms are collected in the MOT from the surrounding background vapour, initially at a rate directly proportional to this background density (inset). The MOT continues to load until near steady-state, with an associated voltage $V_\mathrm{MOT}$ (green square) indicative of the atom number prior to magnetic trapping. Stage (2) depicts the subsequent cooling, optical pumping, and transfer into the magnetic trap, in which the atomic ensemble is held for a variable time $t_\mathrm{MT}$. Initially, a resonant RF field is applied to eject all atoms above a specified energy $E_\mathrm{cut}$ (2.a). During the hold time, collisions with background particles and between trapped \Rb{87} atoms occur (2.b), inducing loss and the redistribution of trapped energies. Finally, a second resonant RF field is applied to eject all atoms above the trap depth $\Emax$ (2.c). Remaining trapped atoms are subsequently recaptured in the MOT by ramping down the field gradient and turning back on the trapping light, as shown in stage (3). The loading curve upon recapture in the MOT is fit to determine $V_\mathrm{MT}$ (red square), indicative of the remaining atom population after the trapping duration.}%
    \label{fig:apparatus}%
    \end{center}
\end{figure*}

For a non-zero trap depth $E_\mathrm{max}$, the loss rate coefficient $\svl = \svt (1- \PQDU)$ deviates from the total collision rate coefficient $\svt$. Collisions with background particles that do not generate loss from the trap redistribute the energies of the trapped particles (heating), in turn modifying the loss rate. To make a proper determination of the relation between the observed loss rate and $\langle \sigma_\mathrm{loss} v \rangle $, it is crucial to quantify the evolution of the energy distribution with hold time $t$, background density $n$, and cut energy $E_\mathrm{cut}$.  The topic of background collision induced ensemble heating has been investigated theoretically \cite{Bjorkholm1988,bali1999quantum}.  Our work here relies on experimental measurements of the energy distribution.

Once the time-dependence of the distribution $\rho(E)$ is known, the change in the trapped atom number $dN$ can be expressed as,
\begin{equation}
	dN = - \left(n \overline{\langle \sigma_\mathrm{loss}(\Emax) v \rangle}(t)+ \Gamma_0\right) N dt 
\end{equation}
where the formal solution can be expressed as an integral over holding time $t$,
\begin{equation}
	\frac{N(t)}{N_0} =\exp\left( - \left[n\int_0^{t}  \overline{\langle \sigma_\mathrm{loss}(\Emax) v \rangle}(t) + \Gamma_0 \right] dt \right).
\end{equation}
Experimentally, we observe the remaining atom number at two times, $t_i$ and $t_k$, given by $N_i$ and $N_k$ respectively. The loss rate is determined by the relative change in the atom number between the two hold times; taking the logarithm of the ratio $N_k/ N_i$, one finds,
\begin{equation}
\begin{split}
	\ln( \frac{N(t_k)}{N(t_i)})&  = -{n_{k}} \int_{0}^{t_k}\overline{\langle \sigma_\mathrm{loss}(\Emax) v \rangle} dt \\
	 &\hspace{1.2em}+ {n_{i}} \int_{0}^{t_i}\overline{\langle \sigma_\mathrm{loss}(\Emax) v \rangle} dt  -\Gamma_0 (t_k - t_i) \\
	  &\simeq  -n \int_{t_i}^{t_k}\overline{\langle \sigma_\mathrm{loss}(\Emax) v \rangle} dt - \Gamma_0(t_k -t_i)
\end{split}
\label{eq:lossrateevolution}
\end{equation}
where we have assumed $n_i$ and $n_k$, the background densities of rubidium at times $t_i$ and $t_k$, respectively, are approximately equal. This is valid for sufficiently short-time scales between measurements at times $t_i$ and $t_k$.  Experimentally, we satisfy this  assumption by operating in a regime where the background density varies over the timescale of hours, whereas the timescale between measurements at $t_i$ and $t_k$ is a few seconds.

To isolate losses due to collisions with background rubidium, we perform loss rate measurements at different background rubidium densities, $n_1$ and $n_2$, where $n_1 \gg n_2$. For the experiments conducted at low background rubidium density, $n_2$, the collision-induced heating rate is negligible over the experimental duration. In principle, the degree of heating due to background collisions with a single species, measured over a single lifetime, is constant for all $n$. However, in this experiment, a variety of species are present, and at low $n$, the relative partial pressure of rubidium is significantly lower than other species in the background, most notably $\mathrm{H}_2$. Owing to the relatively large $U_\mathrm{d}\simeq 21.5$ mK for Rb-$\mathrm{H}_2$ collisions \cite{booth2019universality}, the majority of collisions with background hydrogen result in loss from the trap, thereby lowering the collision-induced heating rate in this regime. Consequently, the ensemble energy distribution, $\rho(E)$, is approximately constant and the temporal average is given by,
\begin{equation}
\int_{t_i}^{t_k} \overline{\left<\sigma_\mathrm{loss}(\Emax) v \right>}_2 dt = \overline{\left<\sigma_\mathrm{loss}(\Emax) v \right>}_2 \left(t_k - t_i \right).
\label{eq:sloss2bar}
\end{equation}
The measured trap loss rate, $\Gamma_2(\Emax)$, can then be expressed as,
%
%
%
\begin{equation}
	\Gamma_2(\Emax) = \Gamma_0 + n_\mathrm{2} \overline{\langle \sigma_\mathrm{loss}(\Emax) v \rangle}_2
\label{eq:gamma2}
\end{equation}
Since the background gas composition is not measured, the trap depth dependence of $\Gamma_0$ is unknown. However, this background remains constant over the experiment duration, allowing us to rearrange Eq.~\ref{eq:gamma2} to solve for $\Gamma_0$.

Repeating these measurements at a higher background Rb density $n_1$, we have,
\begin{eqnarray}
\ln{\left(\frac{N(t_k)}{N (t_i)} \right)_{n_1}}&=&-n_1 \int_{t_i}^{t_k}\overline{\langle \sigma_\mathrm{loss}(\Emax) v \rangle}_1 dt \nonumber \\
& & - \Gamma_0(t_k -t_i) \nonumber \\
&=&-n_1 \int_{t_i}^{t_k}\overline{\langle \sigma_\mathrm{loss} v \rangle}_1 dt \nonumber \\
& & - \left[ \Gamma_2 - n_2 \overline{\left<\sigma_\mathrm{loss}v \right>}_2\right]\left(t_k-t_i \right).
\label{eq:lnn1}
\end{eqnarray}
For this measurement the ensemble energy distribution will change appreciably over the holding time due to a non-negligible heating rate.  Because of the large polarizability ($C_6$) of Rb \cite{derevianko2001, c6atominter, c6collisionspectro} and low peak velocity $v_p$ at room temperature compared to $\mathrm{H}_2$, one expects a correspondingly smaller $U_\mathrm{d}$ for Rb-Rb collisions. Consequently, a higher proportion of the background gas collisions result in heating rather than loss (as compared to the proportion at lower Rb pressures). Thus, for these measurements at  $n_1$, the ensemble energy distribution will change appreciably due to particle collisions. Rearranging Eq.~\ref{eq:lnn1}, one has,
\begin{equation}
\begin{split}
	n_1 \int_{t_i}^{t_k}\overline{\langle \sigma_\mathrm{loss} v \rangle}_1 dt = &\ln{\left(\frac{N(t_i)}{N (t_k)} \right)_{n_1}}\\
	&-\left[\Gamma_2 - n_2 \overline{\langle \sigma_\mathrm{loss} v \rangle}_2 \right] (t_k - t_i)
\end{split}
\label{eq:lnn2}
\end{equation}
As a proxy for the Rb background densities, $n_1$ and $n_2$, we use measurements of the magneto-optical trap (MOT) initial loading rates, $R$, observed during the atom collection stage.  As shown in Ref.~\cite{MOTLoadingRate}, this loading rate is directly proportional to the background particle density.  We write,
\begin{equation}
R_j = \alpha n_j
\label{eq:Rj}
\end{equation}
where $\alpha$ is a constant of proportionality that depends upon the solid angle collected by the optical detection system, the photon-to-voltage conversion efficiency of the detector, and trapping parameters of the MOT \cite{boothphotonscat}.  Substituting $R$ into Eq.~\ref{eq:lnn2}, we obtain,
\begin{equation}
\begin{split}
	\frac{R_1}{\alpha} \int_{t_i}^{t_k}\overline{\langle \sigma_\mathrm{loss} v \rangle}_1 dt = &\ln{\left(\frac{N(t_i)}{N (t_k)} \right)_{n_1}}\\
	&-\left[\Gamma_2 - \frac{R_2}{\alpha} \overline{\langle \sigma_\mathrm{loss} v \rangle}_2 \right] (t_k - t_i)
\end{split}
\label{eq:lnn3}
\end{equation}

We now define the time-integrated loss fraction as,
\begin{equation}
	H_{t_i}^{t_k} = \int_{t_i}^{t_k} \overline{\langle \sigma_\mathrm{loss}(\Emax)v \rangle} dt
\end{equation}
and use this to further rearrange Eq.~\ref{eq:lnn3} arriving at
\begin{equation}
	\frac{1}{\alpha} H_{t_i}^{t_k} = \frac{\ln{\left(N(t_i)/N (t_k) \right)} - \Gamma_2 (t_k - t_i)}{R_1 - \frac{R_2}{H_{t_i}^{t_k}} \overline{\langle\sigma_\mathrm{loss}v \rangle}_2(t_k - t_i)},
\label{eq:Htitk}
\end{equation}
where the unknown quantity $\alpha$ is isolated to one side.  However, $\sigmav$ still appears on both sides of Eq.~\ref{eq:Htitk}, preventing the construction of the RHS from purely experimental measurements \textit{a priori} of the determination of $\sigmav$. 

Therefore, we proceed by using Eq.~\ref{eq:Htitk} as a self-consistency requirement between the fitted quantities, $H_{t_i}^{t_k}$ and $\alpha$, and the experimental measurements ($N(t_j)$, $R_j$, $\Gamma_2$, and $t_k-t_i$). We proceed iteratively: we begin by assuming no heating has occurred and approximate the denominator on the RHS as
\begin{equation}
	R_1 - \frac{R_2}{H_{t_i}^{t_k}} \overline{\langle\sigma_\mathrm{loss}v \rangle}_2(t_k - t_i) \simeq R_1 - R_2.
\end{equation}
Under this approximation, Eq.~\ref{eq:Htitk} becomes,
\begin{equation}
		\frac{\overline{\langle\sigma_\mathrm{loss} v \rangle}_1 (t_k - t_i)}{\alpha} = \frac{\ln{\left(N(t_i)/N (t_k) \right)} - \Gamma_2 (t_k - t_i)}{R_1 - R_2}
\end{equation}
which involves only experimentally determined quantities, meaning an initial estimate of $\sigmav$ and $\alpha$ can be obtained, denoted $\sigmav_0$ and $\alpha_0$ respectively. This expression assumes a constant energy distribution $\rho(E)$ for the entirety of the trapping duration and thereby neglects heating by background collisions. These initial estimates can be used to construct the full expression in Eq.~\ref{eq:Htitk}, now including the measured time-dependence of $\rho(E)$, and subsequent fitting provides an updated estimate for $\sigmav$ and $\alpha$. This process is repeated until convergence of the fitted quantities is achieved, typically within 2-3 iterations. The convergence of the fitted quantities $\sigmav$ and $\alpha$ indicates that the self-consistency requirement posed by Eq.~\ref{eq:Htitk} has been fulfilled, and the final fitted values represent the best estimate for the true values given the experimental measurements, corrected for ensemble heating.

\section{Experiment}
\label{sec:exp}
In our experiment, \Rb{87} atoms are loaded from a background vapour at room temperature into a MOT. Trapping light is generated by grating-stabilized and injection-seeded diode lasers denoted pump (repump) resonant to the $5^2S_{1/2} \rightarrow 5^2 P_{3/2}$ $F=2-3^\prime$ ($1-2^\prime $) $D_2$ transition, detuned by $\delta \simeq -2 \gamma$ ($\delta_\mathrm{repump} = 0$) from resonance, where $\gamma$ is the transition's natural linewidth   \cite{steck2001rubidium}. The magnetic field configuration is a spherical quadrupole with an axial gradient of 27.5 G/cm.

Atomic fluorescence is collected and focused onto a photodetector inducing a voltage proportional ($V$) to the number of atoms present ($N$). To efficiently transfer atoms to the MT, the ensemble is cooled by far-detuning ($\Delta \simeq -6 \gamma $) the pump laser for 50 ms. Next, the pump light is extinguished and the atoms are polarized by a weak uniform field and optically-pumped by a $\sigma^+$ polarized beam resonant to the $F=2-2^\prime$ transition for 5 ms to generate a pure, polarized sample of $\ket{F=2,m_F=2}$ atoms. Residual $\ket{2,1}$ atoms are then ejected by holding the sample at a magnetic field gradient to 27.5 G/cm for which only the $\ket{2,2}$ state is held against gravity. This results in a typical atom transfer efficiency of 40\% of the ensemble loaded into the MT.

The magnetic field is then ramped to a pre-selected trapping gradient and held constant for the hold duration $t$, after which the remaining atoms are recaptured in the MOT. By comparing the MOT fluorescence upon recapture to the fluorescence prior to transfer to the magnetic trap, the relative change in the atom number can be determined,
\begin{equation}
	\frac{N_\mathrm{MT}(t)}{N_\mathrm{MOT}} = \frac{V_\mathrm{MT}(t)}{V_\mathrm{MOT}}.
\end{equation}
This ratiometric measurement minimizes the effects of the shot-to-shot variation of the number of atoms initially loaded into the MT.  A schematic of the apparatus and the experimental sequence, showing an example fluorescence signal, is shown in Fig.~\ref{fig:apparatus}.

\begin{figure}[b]
	\includegraphics[width=0.5\textwidth]{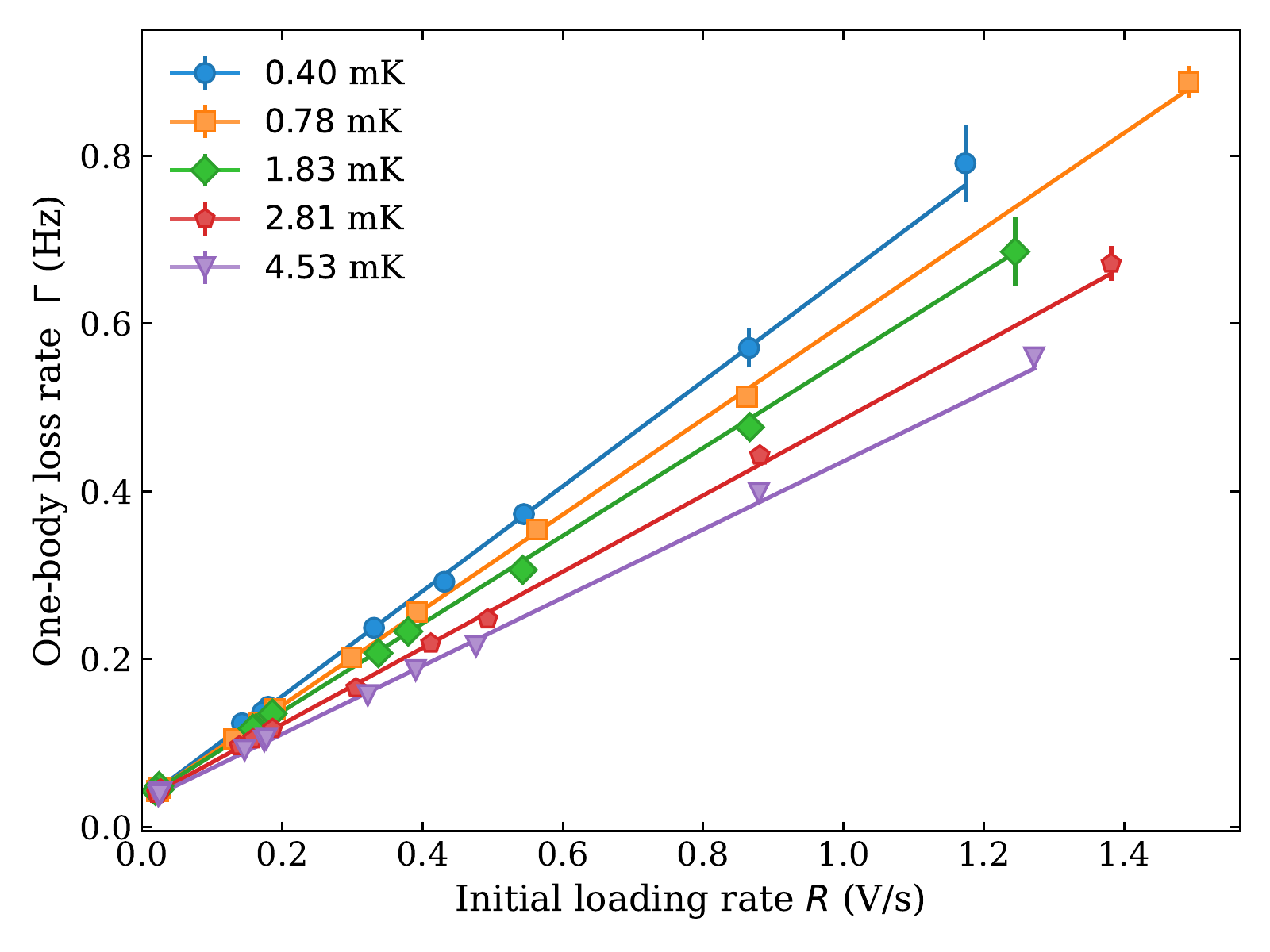}
	\caption{Measured one-body loss rates $\Gamma$ from the magnetic trap as a function of initial loading rate $R$ for varying trap depths $0.40\ \rm{mK} \leq \Emax 
	\leq 4.53$ mK. The loss rate is determined by fitting the decaying atom population to Eq.~\ref{eq:onebody}. The slope varies according to the relation between the background density $n$ and the loss rate at finite trap depth, as per Eq.~\ref{eqn:gloss_g0}. The observed linearity between $\Gamma$ and $R$ indicates that only the background density of rubidium is significantly varying. As $R$ decreases to zero, $\Gamma$ approaches the loss rate induced by collisions with all other background species and Majorana spin-flips.}
	\label{fig:linearityGR}
\end{figure}

The background density of rubidium is varied by energizing a commercial rubidium dispenser for 1-2 minutes, increasing the background density by a factor of 10 or more. During this process, other atomic and molecular species are also released by heating the dispenser, but owing to the relatively slow pumping rate of rubidium from the chamber compared to the other species, the rubidium background density remains elevated while the background density of the other species decreases quickly (on the order of several minutes) to a low, steady equilibrium value. By contrast, the density of rubidium vapor in the test chamber decreases slowly over several days. This assumption of a low and constant background vapour with an elevated rubidium vapour that changes slowly over time is verified by observing the linearity between the initial loading rate of the MOT, $R$, and the loss rate $\Gamma$ of the MT, as shown in Fig. ~\ref{fig:linearityGR}. Since $\Gamma$ scales with the density of all background species, but $R$ is only dependent on the density of background rubidium, linearity between $\Gamma$ and $R$ indicates that only the background density of rubidium is varying significantly during the experiment.
After loading rubidium into the experimental chamber, we wait 1.5 hours before collecting loss rate data to ensure that the rubidium density is only slowly varying over the data collection period.

To perform the energy precut and set the magnetic trap depth, atoms are exposed at the beginning and at the end of the hold time to a RF field that drives transitions between adjacent hyperfine states $\ket{F, m_F} \rightarrow \ket{F, m_F \pm 1}$.  During the exposure, the RF frequency is swept between $\nu_{\min}$ and $\nu_{\max}$.  Atoms with energies that exceed $h \nu_{\min}$ will encounter a resonant RF field and transition into an untrappable state and are thereby ejected from the trap.  When applied at the beginning of the hold time, this procedure empties the trap of atoms above the cut energy $E_{\mathrm{cut}}= h  \nu_{\min}$. When applied at the end of the hold time, a different minimum frequency is chosen to define a precise upper energy bound on the trapped atoms that are recaptured and counted, with the corresponding trap depth given by $\Emax =h  \nu'_{\min}$ (with $\Emax > E_{\mathrm{cut}})$.



\begin{figure}[t]
	\includegraphics[width=0.5\textwidth]{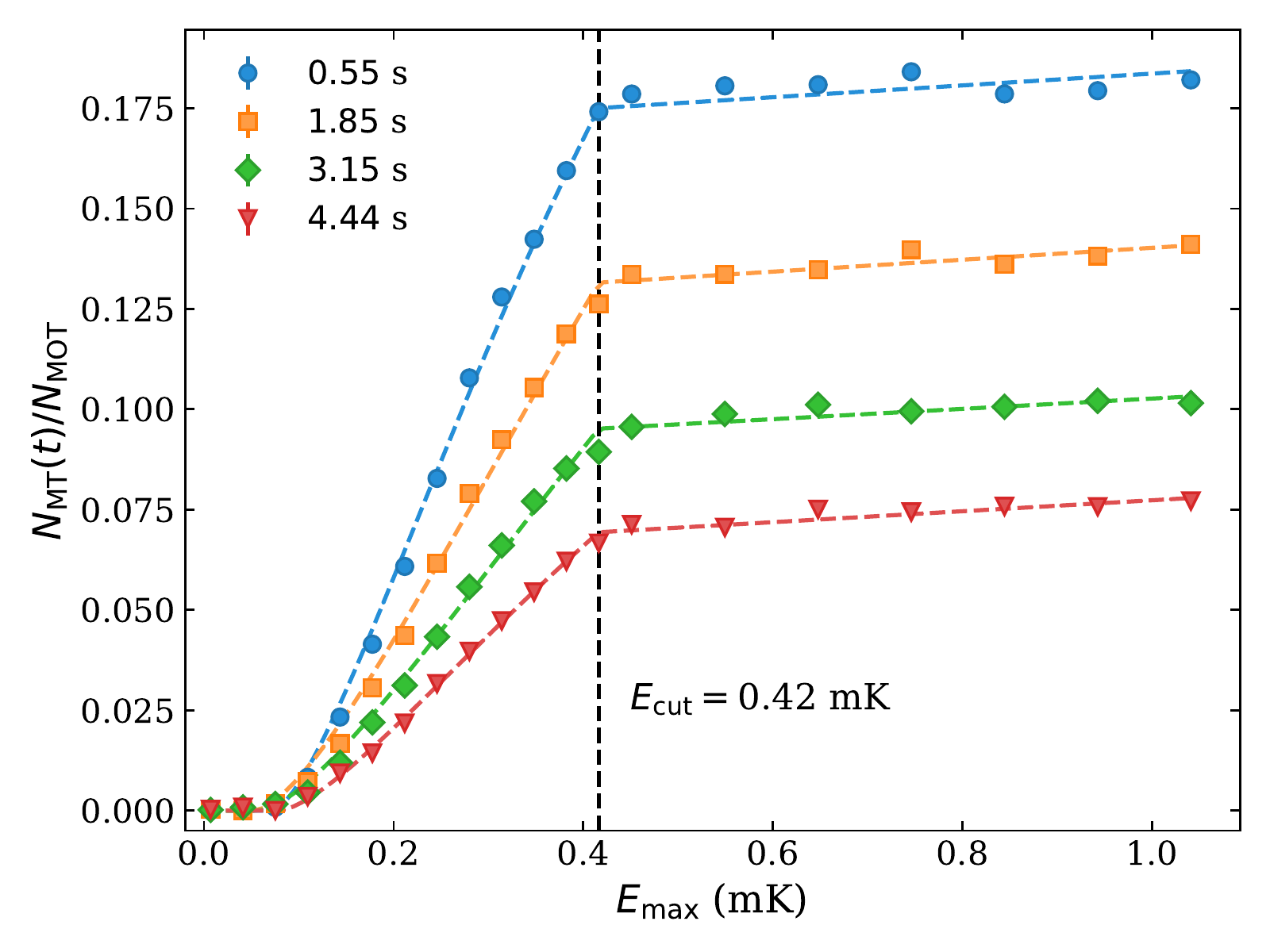}
	\caption{Measured cumulative energy density functions with increasing trapping time, $t$, for $R = 0.504(2)$~V/s, $E_\mathrm{cut} = 0.42$~mK, and an axial trapping gradient of $55$~G/cm. Solid points show the experimentally measured recaptured fraction $N_\mathrm{MT}(t) / N_\mathrm{MOT}$ as a function of trap depth $\Emax$; corresponding dashed lines depict fits to Eq. (\ref{eq:heateddist}). The ensemble is initially truncated at $E_\mathrm{cut}$, shown by the vertical dashed line. As $t$ increases, an increasing fraction of trapped atoms are found above this energy due both intra-trap and background collisions.}
	\label{fig:ensembleevo}
\end{figure}

As described in section \ref{sec:theory}, it is necessary to characterize the evolution of the atoms' energy distribution $\rho(E)$ as both collisional loss and heating occur in the trap.  In order to measure the energy distribution of the atoms in the trap, we apply the RF field to eject all atoms above some energy $E$ and measure the fraction of atoms retained in the trap as a function of this energy.  This measurement is repeated for a variety of energies and provides, as shown in Fig.~\ref{fig:ensembleevo}, a discrete sampling of the cumulative energy distribution (the fraction of atoms below $E$) from which we can extract the energy distribution  $\rho(E)$.

For short trapping durations, we model $\rho(E)$ by a zero-point shifted Maxwell-Boltmann energy distribution,
\begin{equation}
		\rho_\mathrm{MB}(\epsilon, T)= \Theta(E_\mathrm{cut} - \epsilon) \cdot
		 2\sqrt{\frac{\epsilon}{\pi}} \frac{1}{(k_B T)^{3/2}} e^{-\epsilon/ k_B T}
		\label{eq:initialdist}
\end{equation}
where $\epsilon = E - E_\mathrm{min}$, the energy shifted by a fixed amount $E_\mathrm{min}$, below which no atoms are found.  As described above, the ensemble is truncated at $E_\mathrm{cut}$, assuming the initial resonant RF field is applied. As the trapping time increases, we observe a repopulation of states above the initial truncation at $E_\mathrm{cut}$ and an increasing temperature $T$. We describe the heated energy distribution as follows,
\begin{equation}
	\rho(\epsilon, T(t, n)) = \begin{cases}
		\rho_\mathrm{MB}(\epsilon, T(t,n)) & E < E_\mathrm{cut} \\
		m(t, n, E_\mathrm{cut}) & E > E_\mathrm{cut}
	\end{cases}
	\label{eq:heateddist}
\end{equation}
where $m(t,n, E_\mathrm{cut})$ is a uniform distribution indicative of the population above the initial cut energy, as shown in Fig. \ref{fig:ensembleevo}.
 These measurements are repeated with varied parameters $n$, $t$, and $E_\mathrm{cut}$ to fully characterize the dependence of $T$ and $m$. 
 
To determine the loss rate at each trap depth $\Emax$, the relative remaining fraction of the atom population $N_\mathrm{MT}(t) / N_\mathrm{MOT}$ is measured as a function of holding time $t$. We first verify the atom loss is dominated by one-body losses, that is, the relative remaining fraction is well-modelled by a decaying exponential as per Eq.~\ref{eq:onebody}. Experimentally, the minimum holding time is limited by the RF exposure duration required to eject all atoms above the selected precut energy. Following Ref. \cite{Jones1996}, we proceed in subsequent measurements by placing half of the points at a shortest holding time $t_i = 0.55$~s and the remaining points  measurements at $t_k = 1.2 / \Gamma$. Experimentally, we find this sampling scheme provides the most precise measurement of the loss rate given a finite total number of measurements. For this work, we select a total number of 6 measurements for each loss rate, to which the loss rate can be determined with a relative uncertainty less than 1\%. For each of these points, the associated initial loading rate $R$ is also determined, thus providing all the necessary experimental quantities to determine $\sigmav$.



\section{Results}

\begin{figure}[]
	\includegraphics[width=0.5\textwidth]{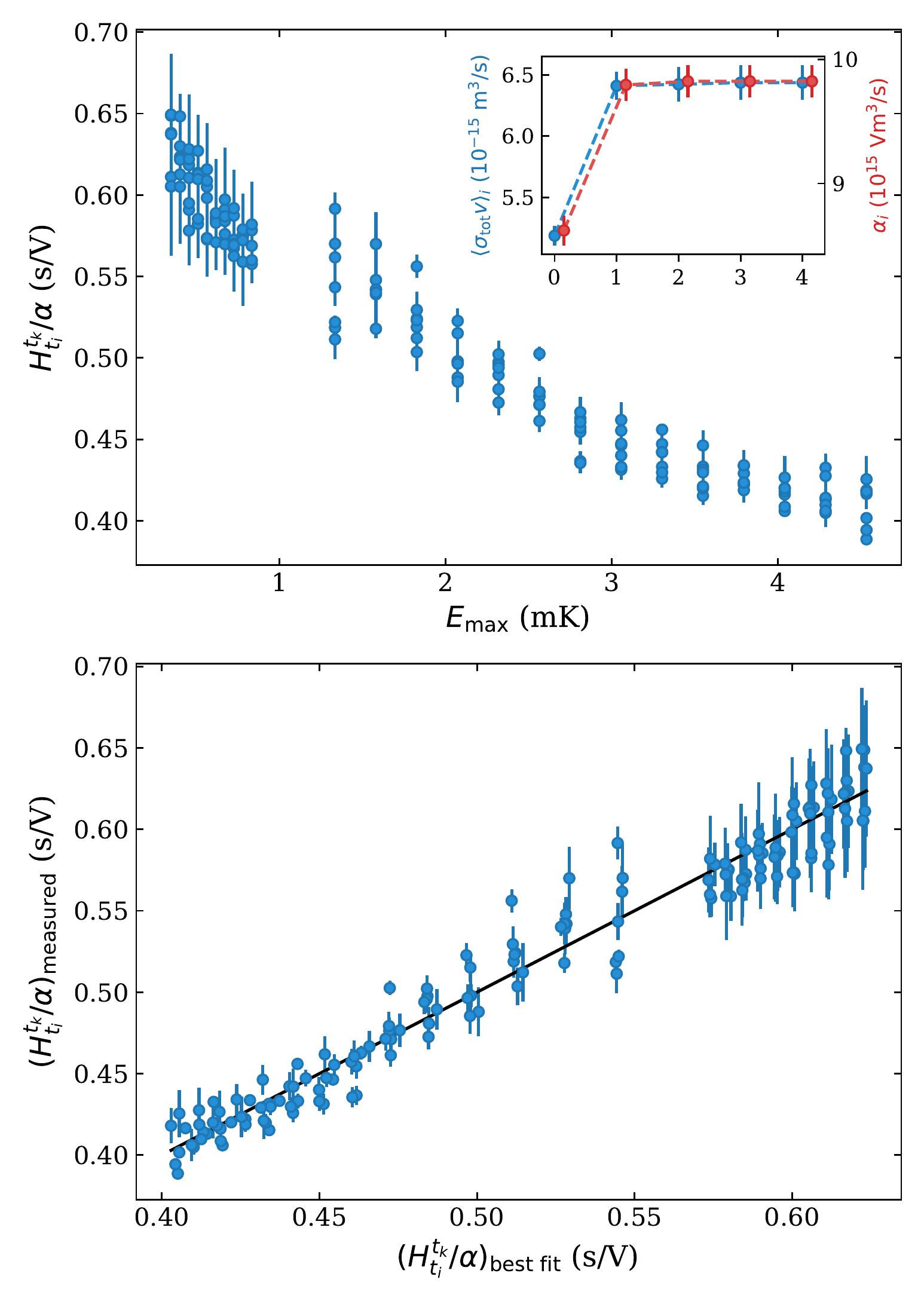}
	\caption{ (Top) Measured time-integrated loss fraction $H_{t_i}^{t_k}/\alpha$ as determined by collisions between trapped and background rubidium as a function of maximum trap depth $\Emax$ following convergence of the iterated fitting process. Each point represents an average over the 6 measurements of the remaining atom number for a particular ensemble energy distribution. Measurements below (above) $\Emax = 1$~mK correspond to an axial trapping gradient of 55 (275)~G/cm.  Inset shows the fitted $\svt_i$ and $\alpha_i$ values as a function of iteration number, for which the values converge within 3 iterations. (Bottom) Measured $H_{t_i}^{t_k}/\alpha$ plotted against the best fit values calculated from the final values of $\svt$ and $\alpha$ and the associated distributions for each point. The solid line, corresponding to the line $y=x$, demonstrates the best fit to the measurements given the associated distributions.}
	\label{fig:Hfit}
\end{figure}

Fig.~\ref{fig:Hfit} shows the experimentally measured time-integrated loss fraction $H_{t_i}^{t_k} / \alpha$ as a function of the maximum trap depth $\Emax$. As the minimum energy $E_\mathrm{min}$ of the trapped ensemble, determined by the misalignment of the MOT and magnetic trap centres \cite{booth2019universality}, scales with increasing trapping gradient, it is difficult to probe small $\Emax$ with a large trapping gradient. We therefore perform measurements at two different axial trapping gradients of 55 and 275 G/cm, enabling us to achieve both shallow and deep trap depths respectively. 

As described in section \ref{sec:theory}, we perform an iterative fitting analysis to account for the redistribution of energies (i.e.~heating) in the trap due to background collisions. Table \ref{tab:fitresults} lists both the initial and final fitting result corresponding to a simultaneous fit across both trapping gradients. As $H_{t_i}^{t_k} / \alpha$ depends on the ensemble energy distribution $\rho(E)$ over the duration in the MT, we include an additional plot of the measured values against those calculated from the final values of $\svt$ and $\alpha$, $(H_{t_i}^{t_k} / \alpha)_\mathrm{best \ fit}$, based the associated distributions for each point. Agreement between these two values is reflected in the grouping of points about the line $y=x$. The variation of the $H_{t_i}^{t_k}/\alpha$ values observed at each $\Emax$ arises from both statistical variations of the determined loss rate (that worsens at lower trap depths where the number of recaptured atoms is smaller) and systematic variations of the loss rate due to differences in the ensemble energy distribution evolution at different background densities.

It is important to note that when the atoms are loaded into the MT, the resulting ensemble energy is significantly altered by the magnetic potential energy associated with the chosen gradient.  As a result, measurements at lower and higher magnetic field gradients reveal different ensemble energy distributions. 
The data presented in Fig.~\ref{fig:Hfit} demonstrates the influence of this distribution change: the data acquired for $\Emax < 0.75$~mK corresponds to a lower magnetic field gradient of 55~G/cm, while those shown for $\Emax > 1.5$~mK were collected at the higher gradient of 275~G/cm. For the colder ensemble---or equivalently, the lower gradient--the average effective trap depth $\Emax - \langle E\rangle$ is larger than that of the hotter ensemble at the higher gradient. As such, the loss rate of the colder ensemble decreases faster as compared to the warmer ensemble at the same trap depth.

The initial fitting value, $\sigmav_0$, corresponds to neglecting the heating of the ensemble and is listed in column 2 of Table~\ref{tab:fitresults}  (`No heating'). This value is systematically  20\%  smaller than the best fit $\sigmav$ value found in column 3. This can be understood from the decrease in the average effective trap depth as the temperature of the ensemble increases, resulting in an increase in the loss rate over time, an effect that is absent when no ensemble heating occurs. The systematic error will be compounded by the fact that the heating rate increases for deeper traps, leading to the observed loss rate appearing to decrease more slowly as a function of trap depth. Consequently, the fit value of $U_\mathrm{d}$ extracted is systematically larger than the true value and hence, yielding the smaller $\sigmav$ value. Accounting for this redistribution, one obtains accordingly a much larger $\sigmav$ value that is closer to the true value (column 3).

We also find clear evidence of two-body intra-trap elastic-collisional losses in the trapped rubidium ensemble. Column 4 of Table~\ref{tab:fitresults} shows the value of $\sigmav$ extracted from the trapped ensembles which were not subjected to the initial RF field to remove those atoms with energies above $E_{\mathrm{cut}}$. For these ensembles, this additional 2-body loss channel is energetically allowed and confounds the extra-trap loss rate measurements, generating the larger value reported here. Note that this value is corrected for ensemble heating using the iterative process reported in this manuscript.

For collisions that are dominated by the long-range van der Waals interaction and subject to the quantum diffractive universality described here, Eq.~\ref{eq:svt_C6} relates $\svt$ to $C_6$. To first order, we observe $C_6 \propto \svt^{\frac{5}{2}}$ making the experimentally determined $C_6$ value a sensitive measure of the validity of the $p_{\rm{QDU}}$ collision model for Rb-Rb collisions. Here we report $\svt=6.44(11)(5) \, \times10^{-15}$~m$^3$/s and $C_6=4688(198)(95) \ E_h a_0^6$ (shown in column 3 of Table~\ref{tab:fitresults}). This is in excellent agreement with theoretical \textit{ab initio} calculations and with previous measurements \cite{derevianko2001, c6atominter, c6collisionspectro}. These results, together with the previous measurements of Rb-N$_2$ collisions \cite{booth2019universality, shen2020realization}, provide evidence of the accuracy of the quantum diffractive collision universality predictions for $\svt$.

While knowledge of the rubidium background pressures $n_1$ and $n_2$ is not needed to extract loss rate coefficients, our analysis provides a measurement of $\alpha$, the proportionality constant between the initial loading rate and the background density of rubidium, from which these may be calculated. At the largest and smallest initial loading rates of $0.014(2)$ V/s and $1.578(7)$ V/s, corresponding to the highest and lowest background densities of rubidium, we find densities of $1.4(2)\times 10^{12}$ $\mathrm{m}^{-3}$ and $1.61(2)\times 10^{14}$ $\mathrm{m}^{-3}$ respectively.  Partial pressures may be readily calculated from the ideal gas law, for which we find $4.3(5)\times 10^{-11}$ Torr and $4.89(6)\times 10^{-9}$ Torr respectively, within expectations given previous studies with similar order of magnitude loss rates \cite{booth2019universality}.




\begin{table}
\caption{\label{tab:fitresults} Experimental values of $\alpha$ and $\sigmav$ determined by the self-consistency requirement posed by Eq. (\ref{eq:Htitk}). The first and second brackets indicates the statistical and systematic error associated with each parameter (see appendix for details). We include previously \textit{ab initio} calculated and experimentally measured dispersion coefficients for comparison \cite{derevianko2001, c6atominter, c6collisionspectro}. Here, $E_h$ is the Hartree energy and $a_0$ is the Bohr radius.}
\begin{tabular}{lccc} \toprule
    {} & {No heating} & {Heating corrected} &  {No precut}\\ \midrule
    $\alpha $ ($10^{15} \frac{\mathrm{V}\mathrm{m}^3}{\mathrm{s}}$) & 8.6(2) & 9.8(2) & 12.5(2) \\
    $\sigmav$ ($10^{-15}\frac{\mathrm{m}^3}{\mathrm{s}}$) & 5.19(8)(4) & 6.44(11)(5) & 7.40(10)(6)\\[0.15em]
    $U_\mathrm{d}$ (mK) & 3.24(5)(3) & 2.61(5)(1) & 2.27(3)(1) \\
    $C_{6,\mathrm{exp}}$ $( E_h a_0^6)$ &  &4688(198)(95) & \\
    \midrule
    \midrule
    $C_{6,\mathrm{theory}}$ $(E_h a_0^6)$ & {} & {4691(23)\cite{derevianko2001}} & {} \\
    \midrule
    $C_{6,\mathrm{exp}}$ $(E_h a_0^6)$ & \multicolumn{3}{c}{4719(30)\cite{c6atominter}, 4700(50)\cite{c6collisionspectro}} \\
\bottomrule
\end{tabular}
\end{table}


\section{Conclusion}

We have presented precision measurements of loss rates of magnetically-trapped \Rb{87} generated by collisions with room temperature rubidium in the surrounding vacuum. We use these measurements, in conjunction with the universal law describing these room temperature collisions, to extract the total velocity-averaged loss rate coefficient $\sigmav$ and the corresponding $C_6$ value for Rb-Rb interactions.

By eliminating two-body intra-trap collisions and accounting for collision-induced ensemble heating, we observe an experimentally measured velocity-averaged total collision rate coefficient $\sigmav = 6.44(11)(5) \times 10^{-15}$ m$^3$/s and a corresponding dispersion coefficient $C_6 = 4688(198)(95)$ $ E_h a_0^6$, in agreement with a previously published theoretical value of $4691(23) E_h a_0^6$ \cite{derevianko2001}. We also find agreement with other experimentally measured $C_6$ values derived from measurements of the Rb ground-state polarizability using atom interferometry \cite{c6atominter} and Feshbach resonance spectroscopy \cite{c6collisionspectro}.


This agreement serves as another experimental verification of quantum diffractive universality for long-range van der Waals interactions and provides independent corroboration that atomic sensors can be used to realize a self-defining, quantum, primary pressure standard.  In addition, this work  demonstrates the importance of accounting for ensemble heating and eliminating two-body intra-trap collisions to minimize systematic errors in determining $\svt$.  For Rb-Rb collisions we find that the value of $\sigmav$ can be underestimated by up to 20\% when the confounding effects of trapped ensemble heating are ignored. It must be noted that this is likely an upper bound on the error as the present work involves collisions between particles of equal mass, maximizing the momentum and energy transfer per collision. This is reflected in the high value of $\svt$, or equivalently, in the low value of $\ud$ characterizing this collision system. Thus, the ensemble heating is significant between trapped and background rubidium atoms at large trap depths. Finally, we observe that if two-body intra-trap collisions are not mitigated, they can lead to an overestimate of $\sigmav$ by up to 15\%. We have shown that this loss channel can be suppressed by removing sensor atoms with energies above half the trap depth, $\Emax/2$. In doing so, elastic two-body losses are energetically forbidden until higher energy states are populated by background collisions or multiple intra-trap collisions.  For the purposes of realizing a self-calibrating primary pressure standard, accounting for and mitigating these confounding effects is essential.

\section{Acknowledgements}
We acknowledge financial support from the Natural Sciences and Engineering Research Council of Canada (NSERC/CRSNG), the Canadian Foundation for Innovation (CFI), and the British Columbia Institute of Technology (BCIT) School of Computing and Academic Studies.  This work was done under the auspices of the Center for Research on Ultra-Cold Systems (CRUCS).  P.S. acknowledges support from the DFG within the GRK 2079/1 program.

\appendix*
\section{Systematic uncertainty}
\label{sec:appendix}

The calculation of the time-integrated loss fraction $H_{t_i}^{t_k}$ requires knowledge of the energy distribution as a function of time, which is calculated based upon the measured temperature $T$, energy offset $E_\mathrm{min}$, and slope $m$ above the cut energy $E_\mathrm{cut}$. To estimate the systematic uncertainty associated with each of these quantities, we analyze the change in the fitted value of $\sigmav$ as each input parameter is varied. Based upon the variances of these measured input quantities, we find they contribute a combined relative systematic uncertainty of $0.80 \%$ on the resultant $\sigmav$ value.

As the derived quantity $\sigmav$ depends on the peak velocity $v_p$ of the surrounding background gas, there is also a systematic shift related to the assumed temperature of the gas. We place a bound of 1 K on the variation of the background temperature, for which the related relative systematic uncertainty is 0.05 \%. Under the assumption that each source of error is independent, we find a total systematic uncertainty of $0.81 \%$ on the determined $\sigmav$ value. 


\bibliography{MAT_RbRb_bib}

\end{document}